# A Comparative Study of ASR Implementations in Resource-Constrained Wireless Sensor Networks for Real-Time Voice Communication

Inaam F., Qutaiba I. Ali
University of Mosul
Iraq

**Abstract:** This paper investigates the challenges and trade-offs associated with implementing Automatic Speech Recognition (ASR) in resource-limited Wireless Sensor Networks (WSNs) for real-time voice communication. We analyze three main architectural approaches: Network Speech Recognition (NSR), Distributed Speech Recognition (DSR), and Embedded Speech Recognition (ESR). Each approach is evaluated based on factors such as bandwidth consumption, processing power requirements, latency, accuracy (Word Error Rate - WER), and adaptability to offline operation. We discuss the advantages and disadvantages of each method, considering the computational and communication limitations of WSN nodes. This comparative study provides insights for selecting the most appropriate ASR implementation strategy based on specific application requirements and resource constraints.

## I. INTRODUCTION

Wireless voice transmission represents an expeditious communication mechanism and therefore unsurprising that it is always wanted in a wide range of emergency scenarios. Generally, in emergency scenarios the time plays a dominant role in the rescue operation. For an efficient communication mechanism, voice can be a significant interfacing tool between the persons in the disaster zone and the rescue center [1]. Typically, voice interface doesn't require visual or physical contact with WSN devices. As a result this feature can speed up the input process and consequently speeding up the rescue operation. However, Real-time voice transmission applications have strict requirements by means of end-to-end delay, data losing and dedicated Bandwidth (B.W) during passing the network. So transmitting audio data from disaster zone toward a rescue center with an acceptable quality within Real-Time constraints is an important issue.

In this paper an alternative solution which depends on Automatic Speech Recognition (ASR) system is proposed, implemented and evaluated. ASR technology provides tools to transform human voice signals into text. Consequently this text can be sent toward a rescue center with less B.W. and power instead of sending the original voice signal. So in this paper we aim to evaluate the performance of a proposed VoWSN based on ASR system for efficiently voice transmission system within Real-Time constraints. The workflow principle of our proposed system is depending on a proposed protocol that transforms system category gradually from network dependent ASR system with a full dictionary and language model (i.e. large vocabulary) to fully embedded ASR system with customized dictionary and language model (i.e. small vocabulary ) specified to closer resemble typical domain. The purpose of optimizing the dictionary (Dic.) and Language Model (LM) is to enhance the ASR system performance in terms of decoding time, Word Error rate (WER) and efficiency. As a result, the transformation could eliminate the network dependency which is the main feature of most state-of-art speech recognizer systems. The objectives of this paper are:

1. To analyze the challenges of implementing ASR systems within the context of resource constrained WSNs.
2. To examine three different ASR architectural approaches: NSR, DSR, and ESR, for WSN integration.
3. To compare the performance of each approach based on key metrics such as WER, latency, bandwidth usage, and power consumption.
4. To evaluate the trade-offs associated with each method, considering the computational limitations of WSN nodes.
5. To provide guidelines for selecting the most suitable ASR implementation strategy based on application needs and resource availability.

## II. LITERATURE REVIEW

The speech recognition process of embedded system is realized by some research, but still has a lot of aspects which needs to be improved [2]. However, a few research endeavors have integrated speech recognition technology into WSN systems and explored general framework practicality. The earliest attempt of implementing speech recognition system on embedded resource constrains systems introduced by S. Phadke, et al.. They combined the aspects of both hardware and software design of implementing a speaker dependent, isolated word speech recognition system. They used modified Mel-scaled Frequency Cepstral Coefficients (MFCC) as feature extraction method and Dynamic Time Warping (DTW) as template matching process [3].

Also, the authors C. Shen, et al., presented the design and implementation of a distributed sensor network application for embedded, isolated-word, real-time speech recognition system. They adopted a parameterized-dataflow-based modelling approach to model the functionalities associated with sensing and processing of acoustic data. The associated embedded software implemented on an off-the-shelf sensor node platform and a TDMA access protocol was developed to manage the wireless channel [4].

In addition, the authors F. Sutton, et al., demonstrated the implementation of prototype architecture for automatic single word speech recognition on resource-constrained embedded devices. The experiments results showed that the prototype achieved a high average detection rate of 96%, while only dissipating 28.5 mW for continuous audio sampling and duty-cycled speechrecognition. Audio signal acquisition was performed by a dedicated audio codec [5].

Moreover, the authors, R. P. Raghava, et al., presented the design and implementation automatic single word speech recognition system on embedded devices. The words which are spell are stored in the operating system of Raspbian OS which is implemented on Raspberry Pi hardware kit of ARM 11 processor [6].

Also, the authors, G. G. tolya, et al., analysed ASR performance on utterances recorded by means of wireless sensors. The sound quality of utterances recorded by such sensors contrasts fundamentally from that of the larger audio data bases usually used for acoustic DNN (Deep

Neural Network) training due to the small microphone installed on these devices. They could accomplish a 5% improvement in terms of relative error reduction [7].

## III. AUTOMATIC SPEECH RECOGNITION BACKGROUND

Speech Recognition (is also known as Automatic Speech Recognition (ASR)) is the process of converting a speech signal to a sequence of words, by means of an algorithm implemented as a computer program [8]. Speech understanding goes one step more, and gleans the meaning of the utterance (The longest segment of speech operated on at one time by an automatic speech recognizer) [9]. Before going into details of speech recognition based WSNs subject, some aspects of automatic speech recognition (ASR) technique must be explained. Figure (1) describes basic processing stages that are part of the ASR system and which will be used for clarifying the classification of ASR system based WSN [10].

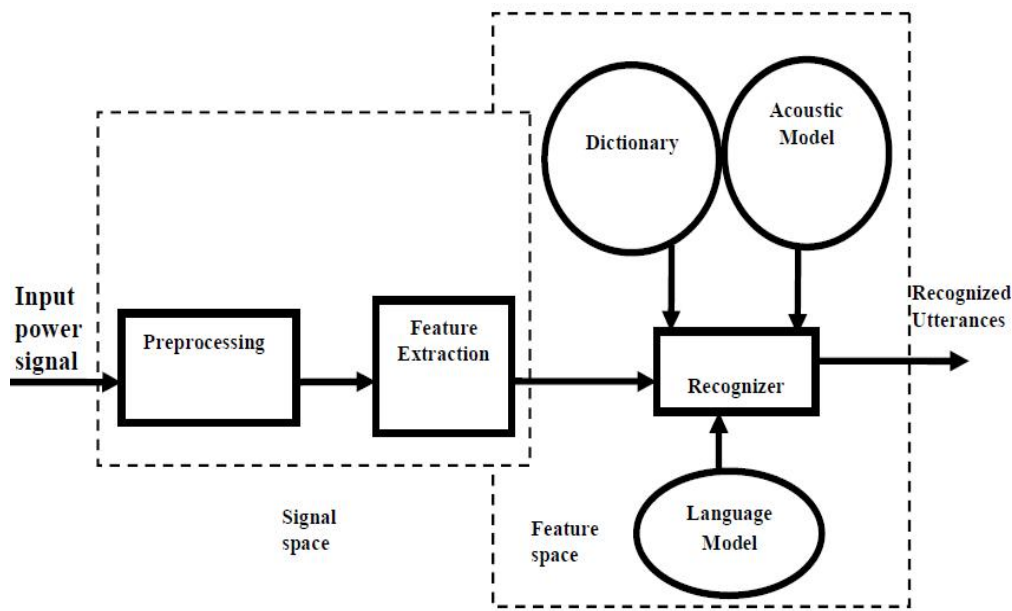

Figure (1). Basic Architecture of ASR system.

Generally, there are two phases in speech recognition system, and the process of features extraction (features extraction is the first stage in ASR process as shown in Figure (4.1)) relevant for classification is common to both phases. The two phases are [11]:

- Training phase as shown in Figure (2): During the training phase, the parameters of the classification model are estimated using a large number of training data.
- Testing phase as shown in Figure (3): During the testing or recognition phase, the features of a test pattern (test speech data) are matched with the trained model of each and every training data. The test pattern is declared to belong to that data whose model matches the test pattern best.

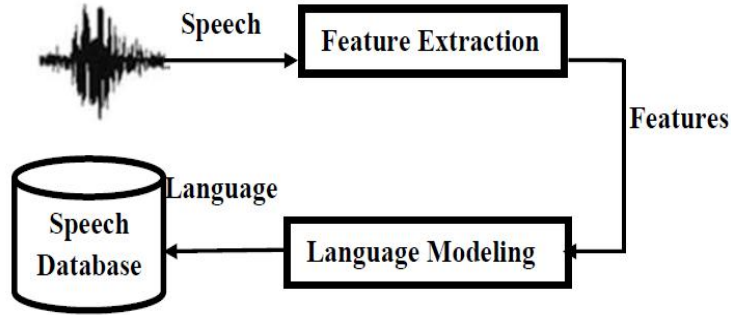
Figure (2). Training Phase Block Diagram.

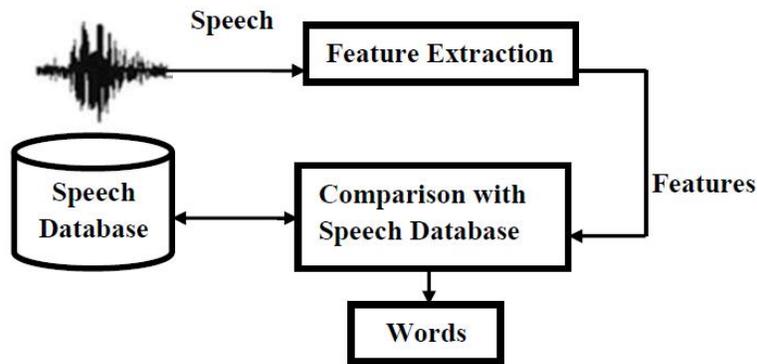
Figure (3). Testing Phase Block Diagram.

### A) Models Used in Speech Recognition

In typical ASR systems three models are used in training and testing phases; an acoustic model, a phonetic dictionary and a language model.

- Acoustic model: The acoustic model is used to translate data from an audio signal to the most probable phones uttered [8]. It describes the sound of words and contains statistical mappings for each phone (or phonemes). Hidden Markov Models (HMMs) are the most powerful parametric model at the acoustic level used in the area of continuous speech recognition and are capable of modeling and matching sequences that have inherent variability in length as well as acoustic characteristics [12]. HMM is using the forward algorithm and Viterbi algorithm [13].
- Phonetic dictionary: The phonetic dictionary maps the relationship between words and its phones. The dictionary may contain several variants of some words, because of differences in the pronunciation. This model can be very large which greatly affects the decoding time of speech recognition. Some systems may for example contain up to several millions of different words.
- Language model: The language model contains statistical information about which words should follow each other in a sentence. This is needed because some words sound very similar to each other taken out of their context. Basically, the role of the language model is to derive the best sentence hypothesis subject to constraints of the language. Because words can have different meanings depending on context, the language model greatly affects the accuracy of a speech recognition system. A static number of previous words are often taken into account when calculating the likelihood of a word. Models using this technique are called N-gram models. By looking at the (*N*-1) previous words the context is taken into account. The language model incorporates various types of linguistic information. The lexicon specifies the sequences of phonemes which form valid words of the language. The syntax describes the rules of combining words to form valid sentences.

### B) Basic ASR Processing Stages

As shown from Figure (1), the speech recognition process can be divided into three consecutive processes. Different speech recognition systems have different implementations of each stage and in between them; the following is an example [14]:

- Pre-processing.
- Feature extraction.
- Classification.

### 1- Pre-processing Process

The input speech signal needs to be processed before extracting features relevant for recognition. The main tasks of pre-processing process are:

- Removing the noise contained in speech signal and filter out any parts of the speech signal that do not contain any speaking [11]. This includes removing silence period that exists at the beginning or end of an audio file as well as parts those contain noise. This is done to counter the fact that the ASR system will assign a probability, even if very low, to any sound-phoneme combination making background noise insert phonemes into the recognition process.
- Re-sample the audio signal to the correct format since the audio signal is required to have a predefined number of channels and a frequency rate (normally 16000 or 8000 KHz which depends on system).

### 2- Feature Extraction Process

Feature extraction process is the most important step in speech recognition and affects greatly the performance of ASR systems. The function of feature extraction phase is extracting features from speech signal and representing them using an appropriate data model of the input signal. This process also known as Front-End analysis[15]. Speech signals are slowly timed varying signals and their characteristics are fairly stationary when examined over a short period of time (5-100 ms). Therefore, in the feature extraction step, acoustic observations are extracted in frames of typically 25 ms. For the acoustic samples in that frame, a multi-dimensional vector is calculated and on that vector a fast Fourier transformation is performed, to transform a function of time, into their frequencies [16]. The following feature extraction techniques are those commonly used in speech recognition [11]:

- Mel-Frequency Cepstrum Coefficients (MFCC)

- Linear Predictive Coding (LPC)
- Linear Prediction Cepstral Coefficients (LPCC)
- Perceptual Linear Prediction (PLP)
- Linear Discriminant Analysis (LDA)
- Discrete Wavelet Transform (DWT)
- Relative Spectral (RASTA-PLP)
- Principal Component analysis (PCA).

However, MFCC has been found to be more robust in the presence of background noise compared to other algorithms [17]. Therefore MFCC is becoming the most evident example of a feature set that is extensively used in speech recognition systems. Figure (4) shows basic block diagram of MFCC feature extraction algorithm [18].

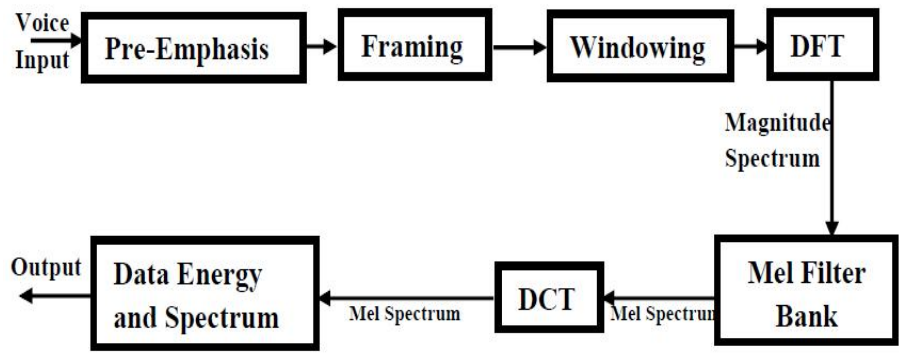

Figure(4). Basic Block Diagram of MFCC Algorithm.

The technique of computing MFCC is based on the short-term analysis, and thus from each frame a MFCC vector is computed. In order to extract the coefficients the speech sample is taken as the input and hamming window is applied to minimize the discontinuities of a signal. Then, DFT will be used to generate the Mel filter bank. MFCC can be computed by using the Formula (1) [13]-[18].

$$\text{Mel}(f) = 2595 \ast \log 10\, (1+f/700) \qquad (1)$$

Where Mel (f): is the Definition of the warping from frequency in Hz to frequency in Mel scale [13].

### 3- Classification Process
In ASR systems there are three approaches for classification process [11]:
- Acoustic Phonetic Approach
- Artificial Intelligence Approach

- Pattern Recognition Approach

The classification of the three approaches including their main algorithms is shown in Figure (5).

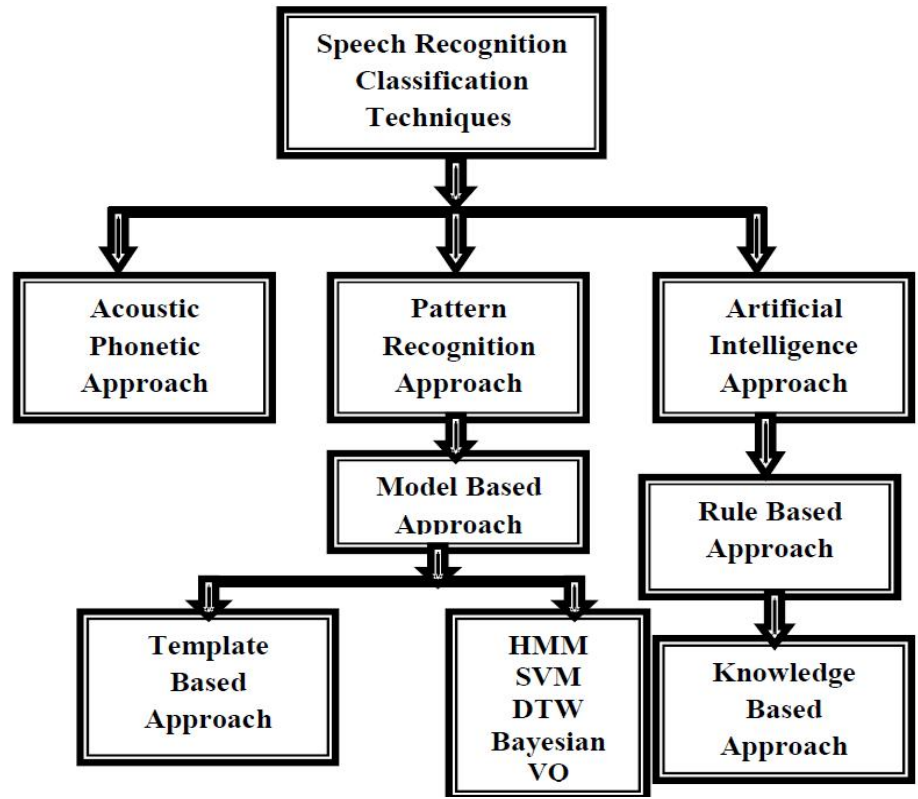

Figure (5). Classification Techniques in Speech Recognition.

- *Acoustic Phonetic Approach:* In this approach the speech recognition relies on finding speech sounds and giving specific labels to these sounds. This is the basis of the acoustic phonetic approach which postulates that "there exist finite, distinctive phonetic units called phonemes and these units are broadly characterized by a set of acoustics properties present in speech" [11].
- *Artificial Intelligence Approach:* This approach attempts to mechanize the recognition process according to the way a person applies its intelligence in visualizing, analyzing and finally making a decision on the measured acoustic features. The artificial intelligence approach is a hybrid of the acoustic phonetic approach and pattern recognition approach. The hybrid concept of both Hidden Markov Model and Artificial Neural Network is also applied in speech recognition. The artificial intelligence approach consists of various methods: Multi-layer Perceptron (MLP), Self-Organizing Map (SOM), Back-propagation Neural Network (BPNN), Time Delay Neural Network (TDNNs).
- *Pattern-Matching Approach:* However, the pattern-matching approach has become the predominant method for most modern speech recognition systems. This approach includes two main steps: pattern training and pattern testing [11]. The main feature of pattern-matching approach is that it uses a well formulated mathematical framework and establishes consistent speech pattern representations for reliable pattern comparison. In the pattern-comparison stage, a direct comparison is made between unknown speeches (the speech to be recognized) with each possible pattern learned in the training stage in order to determine the identity of the unknown data according to the goodness of match of the patterns [8]. This approach contains many techniques such as Hidden Markov Model (HMM), Dynamic Time Wrapping (DTW), Support Vector Machine (SVM), Vector Quantization (VQ) etc as explained in Figure(3). HMM approach is used by most systems to represent the basic units of speech. It is widely used because it is easy, simple and reliable, it can be automatically trained and feasible to use [19].

### C) Mathematical Representation of ASR Process

Generally, the common method used in the modern automatic speech recognition systems is the probabilistic approach which means computing a score for matching spoken words with a speech signal. The aim of the ASR process is to find the most probable sequence of words W = (w1; w2; :::) belonging to a fixed vocabulary given some set of acoustic observations $O$ = ($o$1; $o$2;::: ; $o$T). According to the Bayesian approach when applied to ASR [12] - [20] - [21], the best calculation for the word sequence can be given using Formula (2):

$$W^* = arg_W \max P(W|O) = arg_W \max \frac{P(O|W)P(W)}{P(O)} \quad (2)$$

Since denominator is the same for each candidate sentence W, we can ignore it for the argmax and the best estimation for the word sequence is given by Formula (3):

$$W^* = arg_{W \in L} \max P(O|W) P(W) \quad (3)$$

For generating an output, the speech recognizer has to do the following operations in the sequence that is shown in Figure (6) which shows a block diagram for performing mathematical representation of speech recognition system [22]:

- Extracting the acoustic observations (i.e. features) out of the spoken utterance. In this mathematical representation example, Mel-scaled Frequency Cepstral Coefficients (MFCC) feature extraction algorithm is used.
- Estimating the P(W) - the probability of individual word sequence to happen, regardless acoustic observations.
- Estimating the P($O$|W) - the likelihood that the particular set of features originates from a certain sequence of words.
- Finally, finding the word sequence that delivers the maximum of Formula (3).

The term P (W) is determined by the *language model*. It can be either rule based or of statistical nature. While the likelihoods P ($O$|W) are estimated on most state-of the- art recognizers using Hidden Markov Model (HMM) based *acoustic models*. Here every word wj is composed of a set of acoustic units like phonems, triphones or syllables, i.e. wj = (u1 u2 :::). And every unit uk is modeled by a chain of states sj with associated emission probability density functions p(x|sj ). These densities are usually given by a mixture of diagonal covariance Gaussians using formula 4:

$$p(x|s_j) = \sum_{k=1}^{M} b_{mj} N(x, \mu_{mj}, \Sigma_{mj}) \quad (4)$$

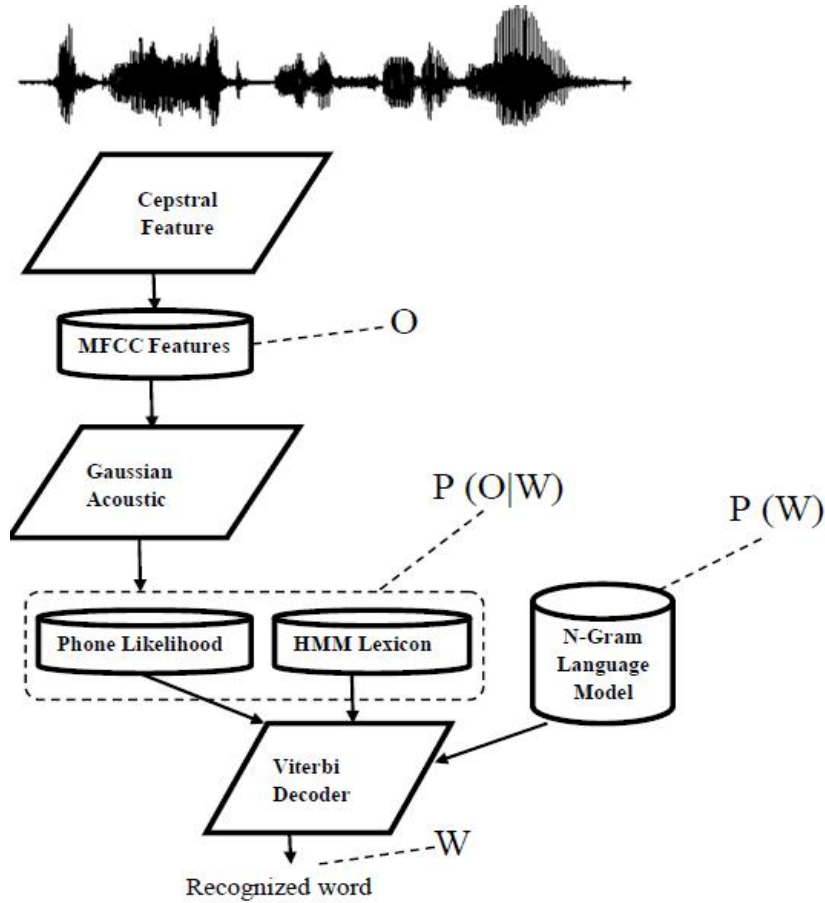

Figure (6). Block Diagram of performing Mathematical Representation of Speech Recognition System.

The computation of the final likelihood P (**O**|W) is performed by combining the state emission likelihoods p(*o*t|sj ) and state transition probabilities. The parameters of acoustic models such as state transition probabilities, means µ$_{mj}$, variances $\Sigma_{mj}$ and weights b$_{mj}$ of Gaussian mixtures are estimated on the training stage and also have to be stored. The total number of Gaussians to be used depends on the design of the recognizer. Finally, armed with both p(*o*t|sj ) and P(W), an effective algorithm is required to explore all HMM states of all words over all word combinations. Viterbi algorithm is usually used to estimate the best word sequence in the relevant lexical tree [20].

Viterbi decoding is a dynamic programming algorithm that searches the state space for the most likely *state sequence* that accounts for the input speech. The state space is constructed by creating word HMM models from its constituent phone or triphone HMM models, and all word HMM models are searched in parallel. Since the state space is huge for even medium vocabulary applications, the beam search heuristic is usually applied to limit the search by pruning out the less likely states. The combination is often simply referred to as *Viterbi beam search*. Generally, Viterbi decoding is a *time-synchronous* search that processes the input speech one *frame* at a time, updating all the states for that frame before moving on to the next. Most systems employ a frame input rate of 100 frames/s [23].

## IV. ASR PERFORMANCE EVALUATION CRITERIA

Typically, the performance evaluation of a typical ASR system covers three important aspects: Speed, average Word Error Rate (WER) and Accuracy. For evaluating the accuracy, the WER must be measured firstly. The meaning of each metric is:

**_Speed:_** This parameter represents the search time on the recognizer and is measured in term of Real Time Factor (RTF) often abbreviated as xRT and calculated as the ratio between the amount of time required to decode an utterance and the length of the utterance using the Formula (5):

$$xRT = \text{Input speech recognition time} / \text{Speech duration} \tag{5}$$

For example, a real-time factor of 0.4 xRT means that each second of audio requires 0.4 seconds to decode (lower RTF means faster decoding) [24].

**_Word error rate_**: Word error rate (WER) is a metric for evaluating the accuracy of a speech recognition system. Calculating WER is a way of measuring the number of errors that occurred during the decoding of an audio signal. Generally, WER is calculated by summarizing the total number of errors in the hypothesis and dividing it by the total number of words in the correct sentence. An error is an incorrect substitution(S) or deletion (D) or insertion (I) of a word which differs the hypothesis from the correct sentence. With this, the word error rate can be calculated using Formula (6) [24]:

$$WER = (I + D + S) / N \tag{6}$$

Where:
**I**: Are the insertion words.
**D**: Are the deletion words.
**S**: Are the substitution words.
**N**: Are the total number of words.
The WER is usually measured in percent. Smaller WER means more efficient ASR system.

**_Accuracy:_** It is almost the same as the word error rate, but it doesn't take insertions into account as described in Formula (7).

$$\text{Accuracy} = (N - D - S) / N \tag{7}$$

## V. DESIGN AND IMPLEMENTATION OF AN ASR SYSTEM BASED EMBEDDED PLATFORMS

Generally, integrating speech recognition into a WSN system produced a new type of applications that have distributed configurations such as smart home automation, health care system, security and so on. Consequently, this research domain has been attract considerable attention for creating innovative energy -saving architectures, algorithms, and protocols to overcome constraints of these embedded systems and meeting the requirements of these applications [25]-[26]. Although many audio monitoring systems utilizing wireless sensor networks are developed over the last few years, but few of these systems are especially targeted at human speech recognition as the main goal of their designs [27]. Figure (7) shows the basic structure of audio wireless sensor network monitoring system.

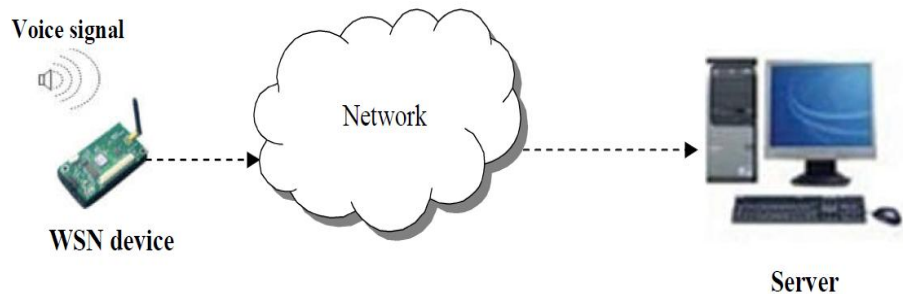

Figure (7). Basic Structure of Audio WSN.

Typically, speech recognition is a high demanding process in terms of computational power and memory storage space. Therefore, ASR based embedded systems need to overcome the capability limitations of WSN devices that defined by:

- Limited computational capability.
- Limited storage space.
- Limited bandwidth.
- Limited energy supply.

As a strategy to resolve the above limitations, a careful compromising between the computation operations and the communication tasks within the system need to be done. This is often achieved by splitting the different functionalities of ASR process between WSN nodes and the Server. In the next paragraph we will discuss strategies for implementing the different categories of ASR system based WSN.

## VI.  CATEGORIES OF ASR BASED WSN SYSTEM

Depending on the decision of where placing the ASR components, three categories of ASR systems based embedded devices can be recognized [10]. The three categories are:

1. Network Speech Recognition (NSR) system.
2. Distributed Speech Recognition (DSR) system.
3. Embedded Speech Recognition (ESR) system.

Each category adopts the client-server architecture and distributes ASR functionalities between the WSN nodes and remote server driven by many factors including device and network resources, ASR components complexity and application demanding.

### 1- Network Speech Recognition (NSR) System

*NSR* system is known to be server-based due to the fact that the overall ASR process is performed by the remote server and the WSN nodes are responsible of capturing, preprocessing and transmitting the speech signal to the server. Figure (8) shows the basic operations distribution between the WSN node and remote server for NSR system. Usually NSR systems use conventional speech coders for the transmission of speech from the embedded device to a recognition server where feature extraction and recognition decoding take place. Google APIs, Alexa are approaches for NSR system.

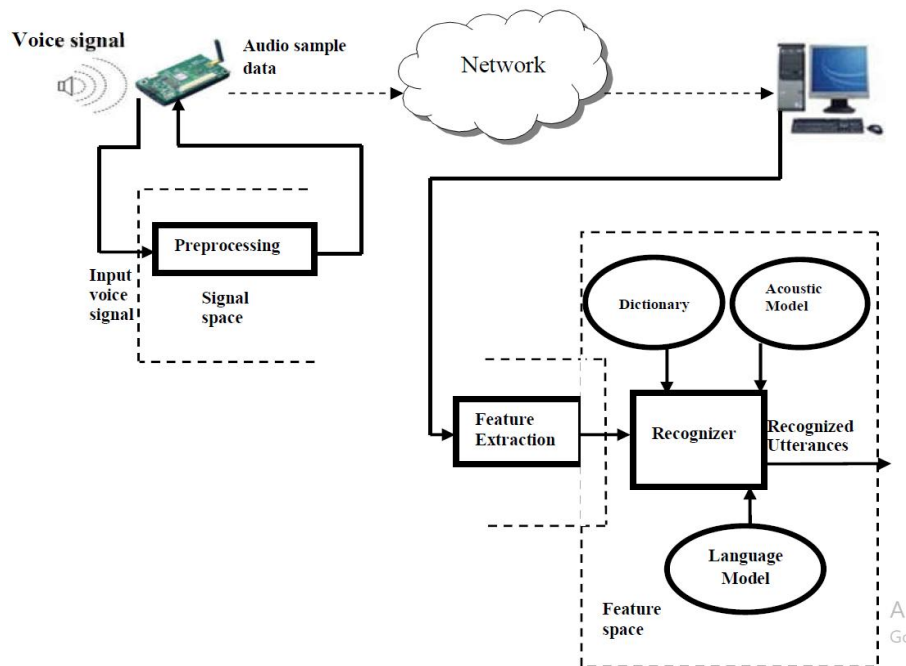

Figure (8) basic operations distribution in NSR system.

Usually, the NSR systems may be preferable when the ASR requires more data from the network than from the microphone and when the ASR computation is a big burden for the embedded device which has limited resources. NSR system can offer many advantages:

1. Supports a wide range of devices in a plug and play fashion.
2. The NSR architecture has low requirement for the client devices (i.e. WSN devices) and can augment these devices, with a very large vocabulary ASR.
3. NSR provides access to the recognizers based on the different grammars or even different languages.
4. The content of the ASR vocabulary frequently might be confidential, consequently restricting its local installation.

5. The NSR system permits an appear-less to the end-user upgrade and alteration of the recognition engine.

On the other side, characteristic drawbacks of the NSR architecture are:
1. Network dependency and error-prone channels.
2. Suffers from transcoding distortion in heterogeneous networks.
3. Recognizer's performance degradation due to using low bit-rate codecs. Low bit-rate speech codec can reduce the recognition performance due to the compression.
4. For Real-Time application, NSR is not an optimal option due to the different sourced of delay which introduced from compression/decompression delay, network delay and recognition delay.

2- ***Distributed Speech Recognition (DSR) System***

Distributed Speech Recognition (DSR) is known to be Client-Server ASR system where the WSN node is responsible of capturing the speech signal, preprocessing it, extracts the acoustical features from the voice and sends them to the Server [15]. When WSN nodes transmit the extracted voice feature values to the Server, further processing will be performed on the Server, including training, and classifying speech features. Figure (9) shows Basic operations splitting in the DSR system.

Due to the restricted resources in term of bandwidth, processing capability and storage capacity, a WSN node cannot sample human voice with a high rate because of the large number of produced samples which need large memory storage. Additionally, transmitting raw sample data will cause too large communication overhead. So implementing an efficient feature extraction algorithm over sensor nodes can reduce audio sample data drastically. The features should aid in discriminating similar sounds and also the number of features should be small so as to reduce computation on WSN node. Advantages of using DSR systems include:

1. The problem of coding and decoding is eliminated since feature extraction process is located in the client and only the features vector (or compressed version of features vector) is needed to be sent to the remote server.
2. The problem of transcoding resulted from passing heterogeneous networks is avoided due to the using of DSR codecs.
3. Robustness against communication channel & acoustic noise.
4. DSR can augment limited resources devices, with a very large vocabulary ASR.
5. Low bandwidth requirement compared to NSR system since the features vector (or compressed version of features vector) is only needed to be transmitted to the server.

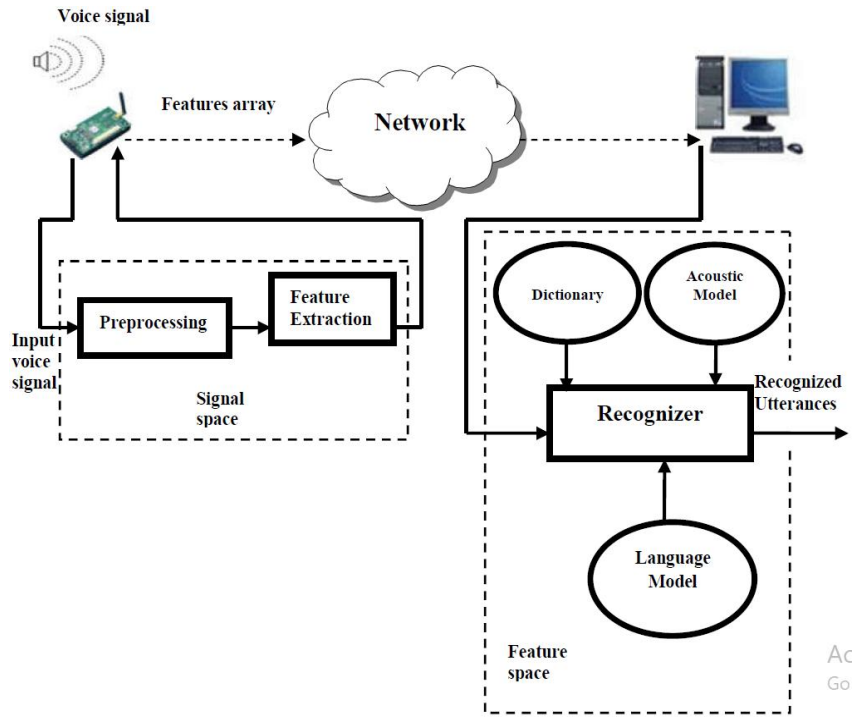

Figure (9). Basic Operations Distribution in *DSR* system.

While drawbacks of using DSR:
1. Front-end analysis (feature extraction process) must be implemented in the WSN. The feature vector analysis routines (and any associated filtering or noise reduction techniques) are the most demanding portion of any speech recognition algorithm, not only because the underlying computations are complex but also because these tasks must be performed in real-time on every speech frame [15].
2. Network dependency and transmission errors.

3- **Embedded Speech Recognition (ESR) System**

The Embedded Speech Recognition (ESR) system is known to be a Client-based ASR system where all ASR processes are performed by the WSN node. Such totally embedded ASR system is independent of network connectivity [20]. However, the requirements of the WSN device are high in terms of computing, memory and power consumption [28]. Pocketsphinx ASR system is an approach for ESR system. Figure (10) shows basic operations splitting in ESR system.

The most important issue for ESR systems is the limited resources of the WSN devices. Therefore, two factors need to be considered when designing the ESR systems:

- The **memory** usage of the underlying algorithms.
- The execution **speed**.

So, to accomplish a "reliable performance of ESR system", the modifications improving both factors should be introduced in every functional block of the ASR system [28] - [20]. One improvement solution of ESR systems is by using fixed-point arithmetic [29]. Using fixed-point arithmetic is the key for low cost and low power consumption, which are important aspects for embedded devices. This solution is adopted by Pocketsphinx [30]. The main advantages of the ESR architecture are:

1. No communication between the remote server and the client is required. Therefore, the ASR system does not rely on the quality of the data transmission and always ready for use.
2. Offline ASR based ESR system can insures privacy since all the ASR process is implemented locally and there is no need to contact the server.
3. For Real-Time application, ESR can be a significant solution due to the eliminations of coding/decoding and network delay resulted from communication with the Server.
4. Since communication consumes power more than other operations, ESR is assumed to be more powerful than NSR and DSR in term of reduced power consumption.
5. Low bandwidth requirement compared to NSR and DSR systems since the text of recognized word is only needed to be transmitted to the server for taking actions.

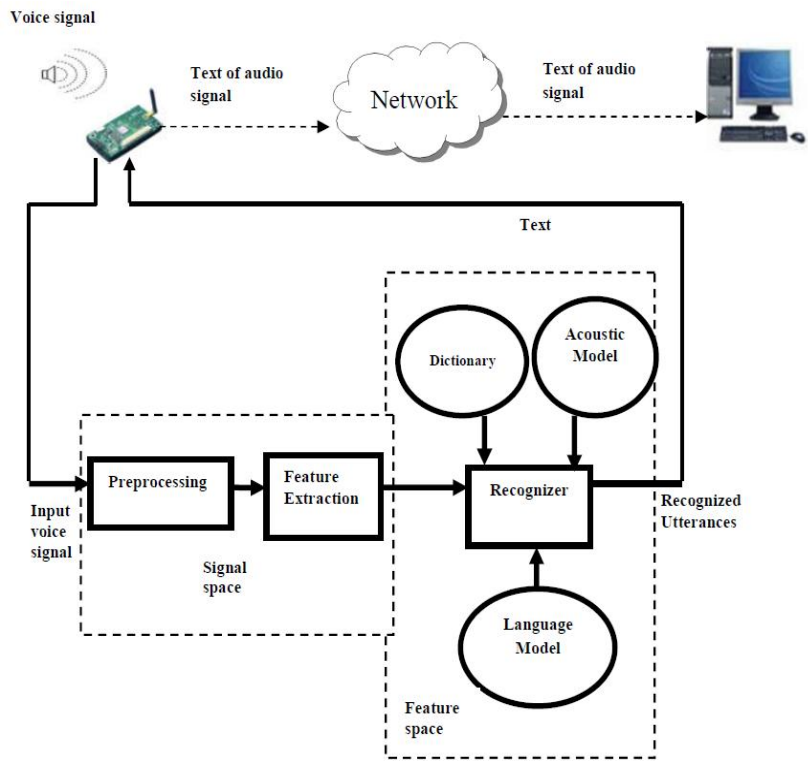

Figure (10). Basic Operations Splitting in *ESR* system.

The following table provides a comprehensive overview of the different ASR implementation strategies, highlighting their strengths and weaknesses. The choice of architecture depends critically on the specific application requirements and the constraints of the WSN deployment. Using this table, researchers and developers can make informed decisions when designing voice enabled WSN applications.

Table (1): a comprehensive overview of the different ASR implementation strategies

| Feature | Network Speech Recognition (NSR) | Distributed Speech Recognition (DSR) | Embedded Speech Recognition (ESR) |
|---|---|---|---|
| Architecture | Server-based; WSN nodes capture and transmit audio. | Client-Server; WSN nodes extract features and transmit them. | Client-based; all ASR processing on the WSN node. |
| Network Dependency | High; Requires continuous network connection. | Moderate; Requires network connection for feature transmission. | Low; Operates independently of the network. |
| Bandwidth Consumption | High; Raw audio data transmission. | Moderate; Feature vector transmission. | Low; Only recognized text transmitted. |
| Processing Power (WSN Node) | Low; Minimal processing on nodes. | Moderate; Feature extraction performed on nodes. | High; All ASR processing on nodes. |
| Latency | High; Network transmission and server-side processing delays. | Moderate; Feature transmission and server-side processing delays. | Low; Minimal processing delay. |
| Vocabulary Size | Large; Can leverage server resources for extensive vocabularies. | Moderate; Limited by network bandwidth and server resources. | Small; Constrained by node resources. |
| Accuracy (WER) | Potentially high; Access to powerful server-side ASR engines. | Moderate; Dependent on feature quality and server-side ASR. | Moderate to Low; Dependent on node resources and optimization techniques. |
| Power Consumption (WSN Node) | Low; Minimal processing on nodes. | Moderate; Feature extraction consumes power. | High; All ASR processing consumes power. |
| Offline Capability | No; Relies on network connection to server. | No; Requires network connection to server. | Yes; Operates independently of the network. |
| Security/Privacy | Potential risks; Voice data transmitted over the network. | Moderate risk; Feature vectors transmitted. | High; Data processed locally. |
| Scalability | High; Easy to add more sensor nodes. | Moderate; Scalability depends on server capacity. | Low; Limited by node resources. |
| Adaptability to dynamic environments | Low; Difficult to adapt to changing network conditions. | Moderate; Can adapt by adjusting feature transmission rate. | High; Can adapt to varying resource availability. |
| Suitable Applications | Applications where network connectivity is reliable and large vocabulary is needed, but latency is not critical (e.g., remote monitoring, voice control in smart homes). | Applications requiring moderate latency and moderate vocabulary size (e.g., distributed voice recognition systems). | Time-critical applications requiring low latency and offline operation (e.g., emergency response, keyword spotting). |

Key Observations & Critical Analysis from this table are listed below:
- Trade-off between Accuracy and Resource Consumption: NSR potentially offers the highest accuracy but consumes the most bandwidth. ESR, while resource-constrained, can achieve acceptable accuracy with optimization. DSR represents a middle ground.
- Network Dependency a Major Factor: NSR and DSR are unsuitable for applications where network connectivity is unreliable or unavailable. ESR excels in offline scenarios.

- Latency Critical for Real-time Systems: ESR offers the lowest latency, making it suitable for real-time applications. NSR suffers from network and server-side processing delays.
- Power Consumption is a Key Constraint in WSNs: ESR's higher power consumption must be carefully considered, especially for battery-powered nodes [31].
- Security and Privacy Implications: NSR raises privacy concerns as voice data is transmitted over the network. ESR, with local processing, offers better privacy.

A demonstrative performance evaluation was conducted to compare NSR, DSR, and ESR implementations, using realistic assumptions for illustrative purposes, see Table (2). A vocabulary size of 50,000, 5,000, and 1,000 words was used for NSR, DSR, and ESR, respectively, with a speech duration of 5 seconds for all. The resulting speed (xRT) was 0.6, 0.8, and 0.2 for NSR, DSR, and ESR, indicating the fastest decoding time for ESR. WER was calculated at 30%, 26%, and 17%, respectively, translating to accuracies of 70%, 74%, and 83%. As expected, ESR achieved the lowest bandwidth consumption (4 kbps) and latency, while NSR demonstrated the highest accuracy but consumed significantly more bandwidth (128 kbps). DSR offered a middle ground in terms of performance. Notably, ESR's power consumption (75 mW) was considerably higher due to local processing on the resource-constrained node, compared to 15 mW and 30 mW for NSR and DSR, respectively. These results highlight the inherent tradeoffs between accuracy, resource usage, and latency among the three architectural approaches.

Table (2): A demonstrative performance evaluation

| Feature | Metric | NSR | DSR | ESR |
| --- | --- | --- | --- | --- |
| Vocabulary Size | Number of words | 50,000 | 5,000 | 1,000 |
| Speech Duration | Seconds | 5 | 5 | 5 |
| Recognition Time | Seconds | 3 | 4 | 1 |
| Speed (xRT) | Equation (5) | 0.6 | 0.8 | 0.2 |
| Insertions (I) | Count | 10 | 8 | 5 |
| Deletions (D) | Count | 5 | 6 | 4 |
| Substitutions (S) | Count | 15 | 12 | 8 |
| Total Words (N) | Count | 100 | 100 | 100 |
| WER (%) | Equation (6) | 30% | 26% | 17% |
| Accuracy (%) | 1 - WER | 70% | 74% | 83% |
| Power Consumption | mW (milliwatts) | 15 | 30 | 75 |
| Bandwidth Consumption | kbps (kilobits per second) | 128 | 32 | 4 |

## CONCLUSIONS:

This study provided a comprehensive comparison of NSR, DSR, and ESR implementations for ASR in WSNs. NSR, while offering flexibility and large vocabularies, is heavily reliant on network connectivity and introduces latency. DSR distributes the computational load, improving robustness but still depends on the network. ESR, though limited
by the resources of individual nodes, offers offline capabilities and low latency, making it attractive for specific time-critical applications. The choice of architecture ultimately depends on the specific application requirements. For applications where network connectivity is reliable and latency is not critical, NSR or DSR may be suitable. However, for applications requiring low latency and offline functionality, ESR, with appropriate optimizations like fixed-point arithmetic and small vocabularies, becomes the preferred choice. Future work should investigate hybrid approaches, combining the strengths of different architectures, and explore advanced techniques for minimizing resource consumption and improving accuracy in challenging environments.